\documentclass[twocolumn,10pt]{article} 

\usepackage[square,numbers,sort&compress,comma]{natbib}

\usepackage{amsmath}
\usepackage{amssymb}
\usepackage{caption}
\usepackage{graphicx}
\usepackage{latexsym}
\usepackage{times}
\usepackage[pagewise]{lineno}
\usepackage{subcaption}
\topmargin - 12pt 
\renewenvironment{abstract}%
              {
               \small
               {\bfseries \abstractname}
               \par
               \vspace{10pt}
              }

\renewcommand\abstractname{Abstract}

\newcommand{\nomenclature}
              [1]
              {
               \bgroup
               \flushleft
               \small\bf
               #1
               \par
               \egroup
              }

\renewcommand{\section}
              [1]
              {
               \bgroup
               \flushleft
               \small\bf
               \refstepcounter{section}
               \arabic{section}. #1
               \par
               \egroup
              }

\renewcommand{\subsection}
              [1]
              {
               \bgroup
               \flushleft
               \small\em
               \refstepcounter{subsection}
               \arabic{section}.
               \arabic{subsection} #1
               \par
               \egroup
              }

\renewcommand{\subsubsection}
              [1]
              {
               \bgroup
               \flushleft
               \small\em
               \refstepcounter{subsubsection}
               \arabic{section}.
               \arabic{subsection}.
               \arabic{subsubsection}. #1
               \par
               \egroup
              }

  \newcommand{\acknowledgement}
              [1]
              {
               \bgroup
               \flushleft
               \small\bf
               #1
               \par
               \egroup
              }

  \newcommand{\sectionbib}
              [1]
              {
               \bgroup
               \flushleft
               \small\bf
               #1
               \par
               \egroup
              }

\begin{document}
\small
\baselineskip 10pt

\title{\LARGE A Physics-Constrained Neural Ordinary Differential Equations Approach for Robust Learning of Stiff Chemical Kinetics}

        
\author{{\large Tadbhagya Kumar$^{a}$, Anuj Kumar$^{a,b}$, Pinaki Pal$^{a,*}$}\\[10pt]
        {\footnotesize \em $^a$Transportation and Power Systems Division, Argonne National Laboratory, Lemont, IL}\\[-5pt]
        {\footnotesize \em $^b$Department of Mechanical Engineering, North Carolina State University, Raleigh, NC}\\[-5pt]}
        
\date{}


\small
\baselineskip 10pt


\twocolumn[\begin{@twocolumnfalse}
\vspace{50pt}
\maketitle
\vspace{40pt}
\rule{\textwidth}{0.5pt}
\begin{abstract} 
The high computational cost associated with solving for detailed chemistry poses a significant challenge for predictive computational fluid dynamics (CFD) simulations of turbulent reacting flows. These models often require solving a system of coupled stiff ordinary differential equations (ODEs), which becomes the computational bottleneck. While deep learning techniques have been experimented with to develop faster surrogate models, they often fail to integrate reliably with CFD solvers. This instability arises because deep learning methods optimize for training error without ensuring compatibility with ODE solvers, leading to accumulation of errors over time. Recently, a first-of-its-kind neural ordinary differential equations (neuralODE$;$ NODE) based approach, known as ChemNODE \cite{ChemNODE}, has been shown as a promising technique to accelerate detailed chemistry computations. In this deep learning framework, the
chemical source terms predicted by the neural networks are integrated during training, and by
computing the required derivatives, the neural network weights are optimized to
minimize the difference between the predicted and ground-truth thermochemical state solutions. In the present work, we extend this NeuralODE framework for stiff chemical kinetics by incorporating mass conservation constraints directly into the loss function during training.  This ensures that the total mass and the elemental species mass are conserved, a critical requirement for reliable downstream integration with CFD solvers. Proof-of-concept studies are performed with the novel physics-constrained neuralODE (PC-NODE) approach for homogeneous autoignition of hydrogen-air mixture
over a range of composition and thermodynamic conditions. It is demonstrated that this enhancement not only improves the physical consistency of the resulting data-driven model with respect to mass conservation criteria but also improves training efficiency. Lastly, \textit{a posteriori} studies are performed wherein the trained PC-NODE model is coupled with a 3D CFD solver for computing the chemical source terms. PC-NODE is shown to be more accurate relative to the purely data-driven neuralODE approach. Moreover, PC-NODE also exhibits robustness and generalizability to unseen initial conditions from within (interpolative capability) as well as outside (extrapolative capability) the training regime. 

\end{abstract}
\vspace{10pt}
\parbox{1.0\textwidth}{\footnotesize {\em Keywords:} NeuralODEs; Physics-informed machine learning; Deep learning; Chemical kinetics; Combustion CFD}
\rule{\textwidth}{0.5pt}
\vspace{10pt}
\end{@twocolumnfalse}]

\clearpage


\section{Introduction\label{sec:introduction}} \addvspace{10pt}

Computational fluid dynamics (CFD) modeling of turbulent combustion remains computationally demanding, which is attributed to the complex interaction of multiple physical phenomena, such as flow turbulence, heat transfer and chemical kinetics, spanning a wide range of spatio-temporal scales. Of these, modeling of detailed chemical kinetics presents the principal bottleneck. The associated kinetics is governed by a stiff system of coupled ordinary differential equations (ODEs) characterized by a large range in magnitude of eigenvalues of the corresponding Jacobian \cite{SHAMPINE1993279}. In addition, the dimensionality of the ODE system increases exponentially as reaction mechanisms get larger (in terms of the number of species and chemical reactions) \cite{lu2009toward}.

In order to make the computations tractable, mechanism reduction is performed, either through the elimination of certain reactions and species \cite{LU20051333, JONES2005223, SUN2010}  or, through analysis of the impact of timescales on the global reactions \cite{VALORANI200629, MAAS1992239}. However, the reduced mechanisms often lead to less reliable description of chemical kinetics compared to the original detailed chemistry. 

Recently, several Machine Learning (ML) approaches have been employed to emulate chemical kinetics and accelerate the associated chemistry integration. 
Some of these approaches reduce the dimensionality of the reaction system by linearly projecting it on an appropriate basis identified typically through Principal Component Analysis (PCA). ML methods are, then, employed for the regression of associated source terms and transport coefficients in the reduced space ~\cite{ope2017, KUMAR2023112903}. On the other hand, data-driven models in the form of feedforward neural networks (FNNs) have also been used to learn multidimensional flamelet lookup tables with various control variables, thereby reducing the associated memory footprint \cite{ZHANG2020100021, OWOYELE20215889, ope2020}. Furthermore, numerous studies have employed artificial neural networks (ANNs) to streamline the computation of chemical source terms \cite{CHRISTO199643, BLASCO199838, SEN2010566, ranade2019, WAN2020119}. Despite the potential benefits of employing ML methods, there are challenges. When coupled with a numerical solver, predicted solutions may diverge or become unstable. The nonlinearity of the combustion process means even minor predictive errors can escalate into significant discrepancies in temporal evolution of the thermochemical state due to error accumulation.


Recently, in an effort to alleviate these challenges, deep learning models based on neural ordinary differential equations (NeuralODEs) \cite{chen2018neural} have been proposed for chemical kinetic calculations \cite{ChemNODE, kim2021}. A first-of-its-kind NODE approach, known as ChemNODE, proposed by Owoyele and Pal \cite{ChemNODE}, obtains the solution vector through a stiff ODE solver operating on an ODE system parameterized by a neural network that outputs the chemical source terms. The neuralODE framework ensures that the obtained solution vectors, even after a long-time horizon, remain adherent to the original solution trajectory. The ChemNODE approach, in addition to the computational savings, accurately predicted the chemical kinetic behavior of homogeneous autoignition of a hydrogen-air mixture.

A major limitation of ML models applied to modeling of physical systems is that they do not inherently incorporate the physical boundaries set by the governing equations and conservation laws. Minor errors associated with conservation laws can accumulate and adversely impact the accuracy of a flow simulation, thereby hindering the integration of ML surrogate models with multidimensional CFD solvers. 

More recently, the practice of combining domain science-specific constraints to embed physics into neural network training has gained prominence \cite{raissi2019physics}. This is achieved through the regularization of the loss function, allowing for integration of physical laws into the learning process. Such an approach significantly enhances the model's ability to handle complex multiphysics problems, offering improved accuracy and generalization, especially in scenarios where traditional methods struggle with noisy data and high-dimensional problems governed by parameterized partial differential equations.

In the present work, the neuralODE framework for chemical kinetics \cite{ChemNODE} is enhanced by incorporating mass conservation constraints in the form of elemental mass conservation into the loss function during ML training, similar to PINNs \cite{raissi2019physics}. It is demonstrated that these modifications to the original approach not only improve the consistency of the resulting data-driven model to the physical laws but also make the training process computationally efficient and predictions from the neural network more robust. 

The rest of the paper is organized as follows. The neuralODE framework is first outlined along with the details of the physics-constrained formulation and training methodology in sections 2 and 3, respectively. Results from \textit{a posteriori} proof-of-concept studies are discussed in section 4. The paper finishes with concluding remarks in Section 5.

\section{Physics-constrained neuralODE (PC-NODE) framework for stiff chemical kinetics\label{sec:sections}} \addvspace{10pt}

For an unsteady chemically reacting system (with no diffusive and advective transport), the temporal evolution of reactive scalars (species) can be defined by:
\begin{equation}\label{Arrhenius}
    \frac{dY_k}{dt} = \frac{\dot{\omega}_k}{\rho}, ~~ k = 1,2,3,...,N_s
\end{equation}
where $Y_k$ is the mass fraction of species $k$, $\dot{\omega}_k$ is the corresponding chemical source term computed using law of mass action, $\rho$ is the density, $N_s$ is the number of chemical species. The temporal evolution of temperature is also governed by an ODE similar to Eq. \eqref{Arrhenius}. To calculate these source terms, one needs to account for several elementary reactions involving production and consumption of multiple species. As the chemical mechanism becomes larger, the number of intermediate reactions also increases when detailed kinetics is taken into account \cite{lu2009toward}. This leads to extreme computational costs, since all chemical time scales must be fully resolved. 

In the ChemNODE \cite{ChemNODE} framework, the expensive physics-based computation of chemical source terms is replaced by a neural network, which can be described as:
\begin{equation}\label{ChemNODE}
    \frac{d \bold{\Phi}}{dt} = f(\bold{\Phi},t; \bold{\Theta})
\end{equation}
where $\bold{\Phi}$ is the vector of thermochemical state (temperature and species mass-fractions, $\bold{\Phi} = [T, Y_1, Y_2,...., Y_{N_s}]$), and $f(\bold{\Phi},t; \bold{\Theta})$ is an FNN parameterized by weights $\bold{\Theta}$. The data-driven learning process is posed as an optimization problem of determining the optimal FNN parameters that minimize the mean-squared error (MSE) loss function:

\begin{equation}\label{MSELoss}
    L_{MSE} = \frac{1}{N(N_s+1)} \sum_{j=1}^{N} \sum_{i=1}^{N_s + 1} \left( \frac{\Phi_{i,j} - \hat{\Phi}_{i,j}}{\Phi_{i,max} - \Phi_{i,min}} \right)^2
\end{equation}
where $\Phi$ is the ground truth thermochemical state obtained by integrating Eq. $\eqref{Arrhenius}$, and $\hat{\Phi}$ is the predicted state at different time instants during the neuralODE integration (Eq. $\eqref{ChemNODE}$). $\bold{\Phi}_{max}$ and $\bold{\Phi}_{min}$ are the maximum and minimum scalar values derived from the training data, so that the loss contributions from different species are equally weighted.  

In this case, the system is required to adhere to the law of conservation of mass to be physically consistent. To this end, the loss function of neuralODE training is modified to include mass conservation in the form of elemental mass conservation constraints, as follows: 
\begin{equation}\label{pinnloss}
    L_{PC-NODE} = L_{MSE} + \sum_{j=1}^{N_{ele}-1}\lambda_j L_{ele-j}
\end{equation}
where $L_{ele-j}$ incorporates the loss associated with mass conservation of element $j$ (in the chemical system with a total of $N_{ele}$ elements). For a chemical mechanism with total of $N_s$ reactive scalars (species), this term can be written as:
\begin{equation}\label{ele_loss}
\begin{split}
    L_{ele-j} & = \sum_{i=1}^{N} \frac{1}{N} e_{ij}^2
\end{split}
\end{equation}
and, 
\begin{equation}
e_{ij} =  log \left( 1 + \left| \sum_{k=1}^{N_s}  \frac{N_j^k W_j}{W_k} \left( Y_{k,i} - \hat{Y}_{k,i} \right) \right| \right) 
\end{equation}
where $W_j$ is the atomic mass of the element $j$, $N_j^k$ is the number of atoms of element $j$ in $k^{th}$ species, $W_k$ is the molecular weight of $k^{th}$ species, and $N$ is the number of training data points. 
%
In the above formulation of the elemental mass constraints, the logarithmic form is used because the input scalars are transformed to log basis. The neuralODEs trained with loss function from Eq. \eqref{pinnloss} are called physics-constrained neuralODE (PC-NODE) in the following sections.


\section{Proof-of-concept study}\addvspace{10pt}
In reacting CFD simulations, it is a common approach to decouple the chemistry from transport using operator splitting. The chemistry is solved independently from advective and diffusive transport within each computational grid cell considered as a homogeneous reactor. In this work, a canonical autoigniting hydrogen-air homogeneous reactor at constant pressure of 1 atm is considered. The detailed chemical mechanism \cite{o2004comprehensive} consists of 9 species ($H_2, O_2, O, OH, H_2O, H, HO_2, H_2O_2, N_2$) and 19 chemical reactions, without $NO_x$ chemistry. 

The training data is generated using Cantera \cite{goodwin2009cantera}, which solves the coupled ODE system (Eq. \eqref{Arrhenius}) with detailed chemistry. The initial conditions chosen for training data generation consist of 5 equispaced initial temperatures in $T_i = [1000$ K, 1200 K$]$ and 6 equivalence ratios in $\phi = [0.5,1.5]$ for a total of 30 initial conditions ($T_i, \phi$). Each of these initial conditions is integrated to equillibrium, and the solution is saved at 50 points in time. A validation dataset consisting of 20 initial conditions is also generated for $T_i = [1025$ K, 1075 K, 1125 K, 1175 K$]$ and $\phi = [0.6, 0.8, 1.0, 1.2, 1.4]$. The neuralODE based chemical kinetics model is initialized with the same training initial conditions as the physics-based mechanism. A two-hidden layer FNN with 48 neurons in each hidden layer is used as the surrogate model ($f(\bold{\Phi},t;\bold{\Theta})$), with hyperbolic tangent (tanh) activation function. The output layer is a dense linear layer. The training framework is implemented in Julia programming language.

The input to the neural network is a vector $\bold{\Phi}$ containing temperature and species mass fractions (except of nitrogen ($N_2$) since it acts as an inert gas and does not vary with time) in the logarithmic space \cite{ChemNODE, kim2021}, and the output is a vector containing chemical source terms for the input vector. The forward pass through this system requires integrating the ODEs, for which an A-L 4th order ESDIRK method from Julia’s DifferentialEquations.jl library \cite{rackauckas2017differentialequations} is used. To calculate the gradients for parameter update, forward mode automatic differentiation is employed. In this study, a second order Levenberg-Marquardt (LM) optimizer is used, wherein the neural network weights are updated according to:
\begin{equation}
\bold{\Theta}^{n+1} = \bold{\Theta}^n - (\textbf{J}^T\textbf{J} + \nu \textbf{I})^{-1} \textbf{J}\textbf{e}  
\end{equation}
where $\textbf{J} = \frac{\partial \textbf{e}}{\partial \bold{\Theta} }$, is the Jacobian matrix, $\textbf{e} = \bold{\Phi} - \hat{\bold{\Phi}}$ is the error vector, and $\nu$ is a damping constant set to an initial value of 1000, and increased as the loss value saturates.

\section{Results and Discussion}\addvspace{10pt}
\subsection{Cantera-NODE a posteriori studies}\addvspace{10pt}
In this section, results from the neuralODE trained with MSE loss (Eq. \eqref{MSELoss}) and the physics-constrained loss (Eq. \eqref{pinnloss}) are presented. To train the PC-NODE, the hydrogen (H) and oxygen (O) based elemental mass conservation terms are considered for the hydrogen-air chemistry in Eq. \eqref{ele_loss} with $\lambda_H = \lambda_O = 3$. 

\begin{figure}[!h]
    \includegraphics[width=0.45\textwidth]{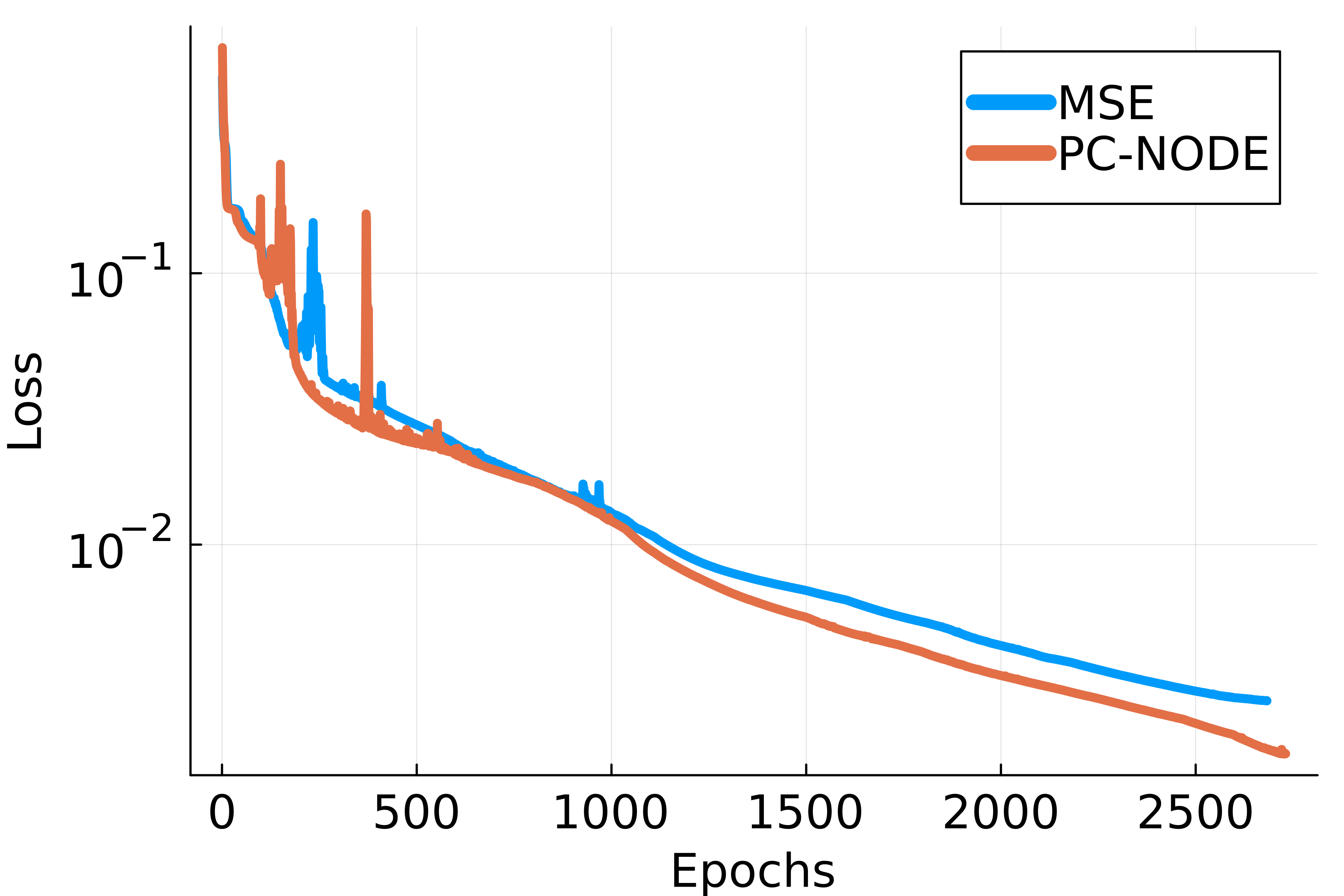}
    \label{loss_plot}
    \caption{Mean squared error (computed between ground truth and predictions) evolution for the two cases trained with MSE loss  function (Eq. \eqref{MSELoss}) and PC-NODE loss function (Eq. \eqref{pinnloss}).}
    \label{fig::logloss}
\end{figure}

Figure \ref{fig::logloss} compares the training loss decay (computed between ground truth and predictions) for neuralODEs trained with MSE loss and with additional elemental mass constraints (PC-NODE). It can be seen that incorporating the elemental mass conservation constraints in the loss function decays it to a lower value and makes the training more efficient. 

\begin{figure*}
     \begin{subfigure}[b]{\textwidth}
         \centering
         \includegraphics[width=0.32\textwidth]{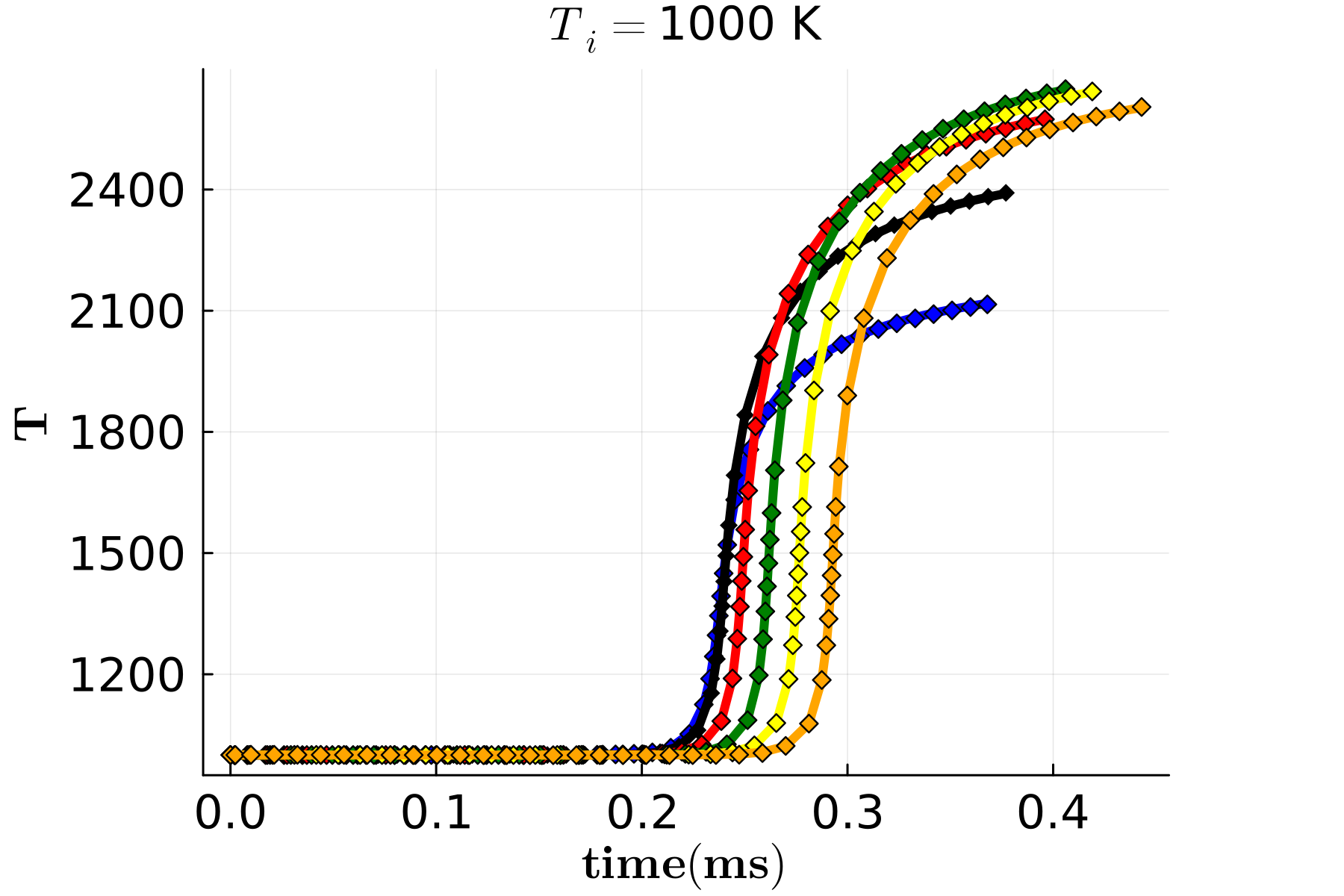}
         \centering
         \includegraphics[width=0.32\textwidth]{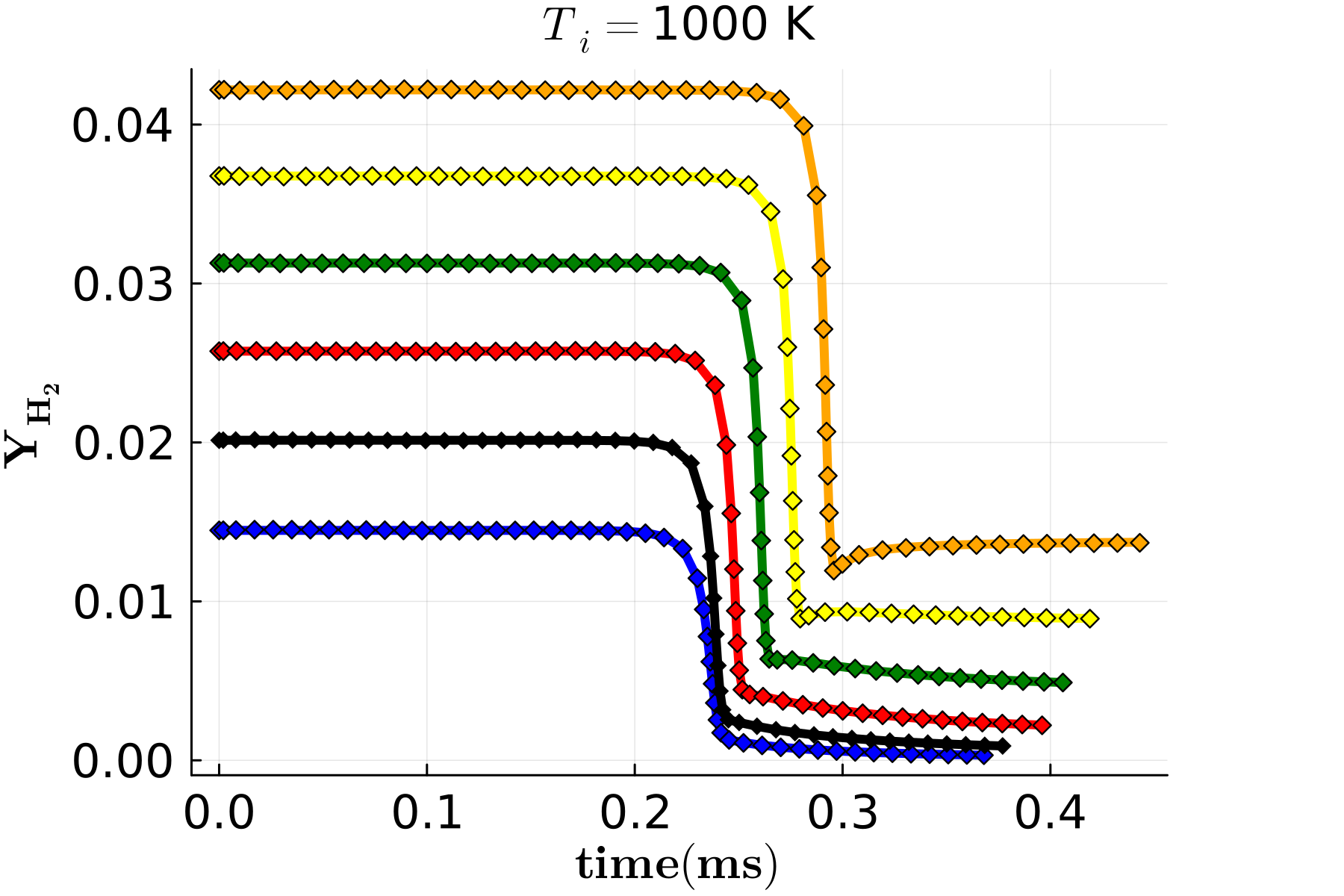}
         \centering
         \centering
         \includegraphics[width=0.32\textwidth]{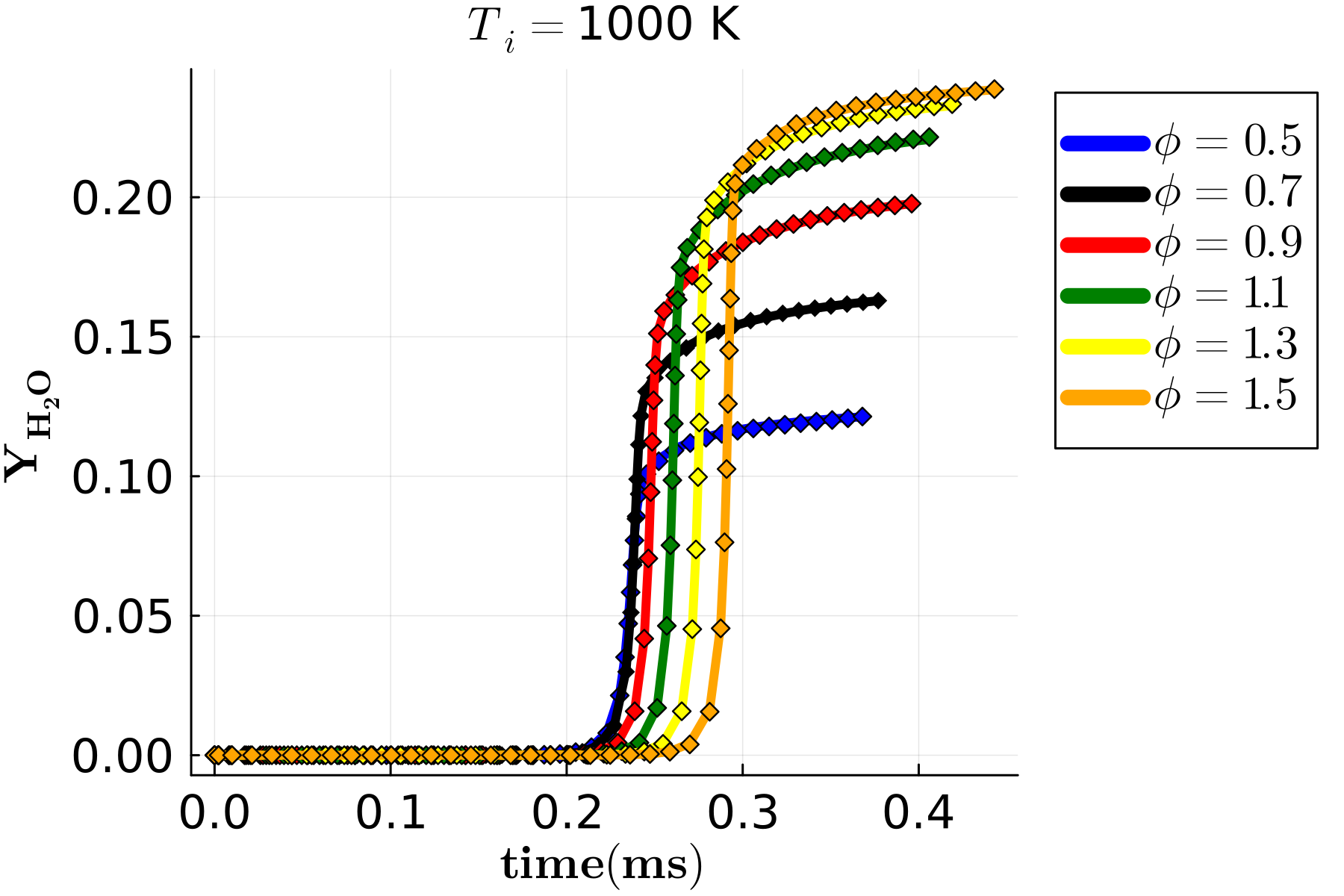}
         \caption{}
     \end{subfigure} \ \
     
    \begin{subfigure}[b]{\textwidth}
         \centering
         \includegraphics[width=0.32\textwidth]{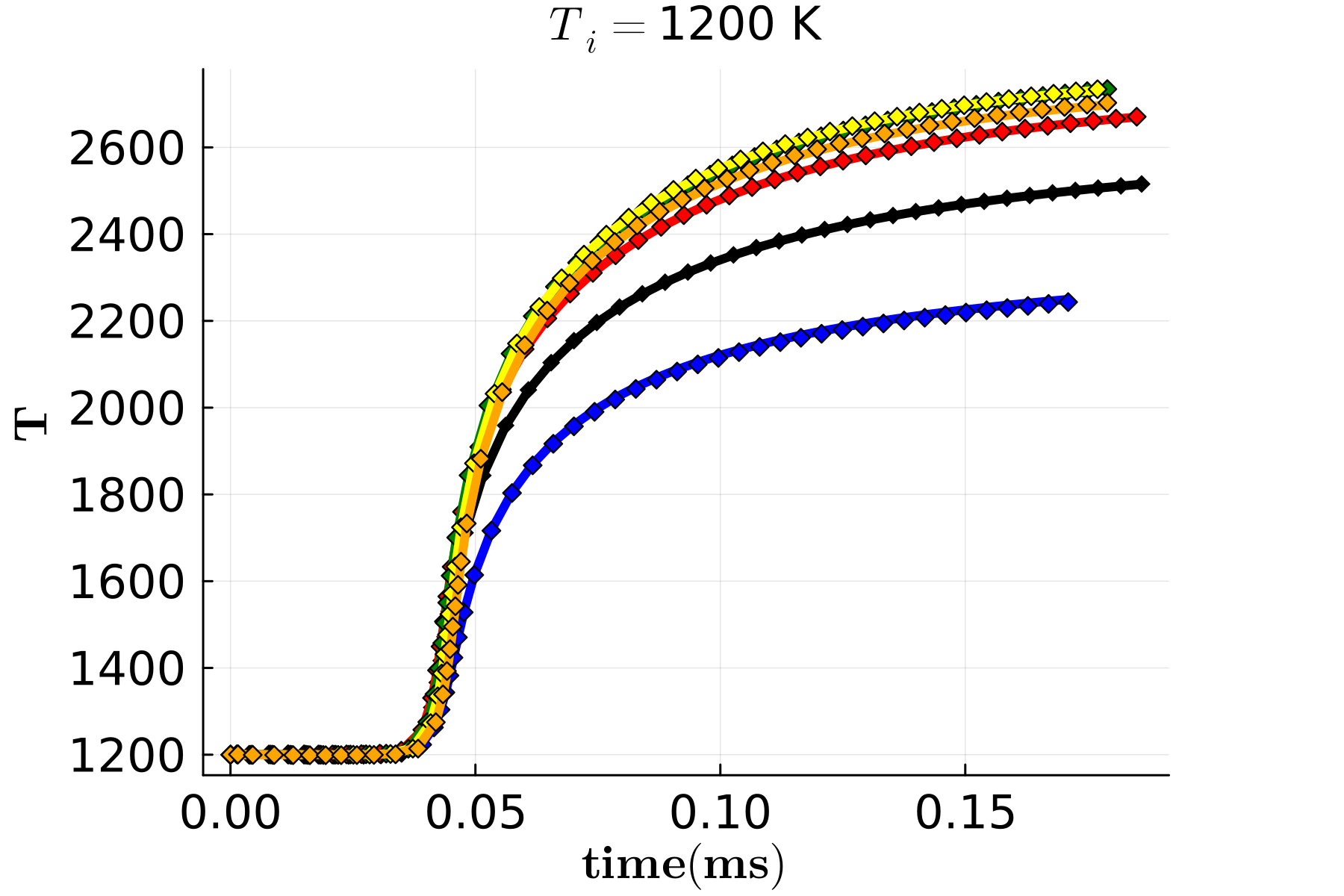}
         \centering
         \includegraphics[width=0.32\textwidth]{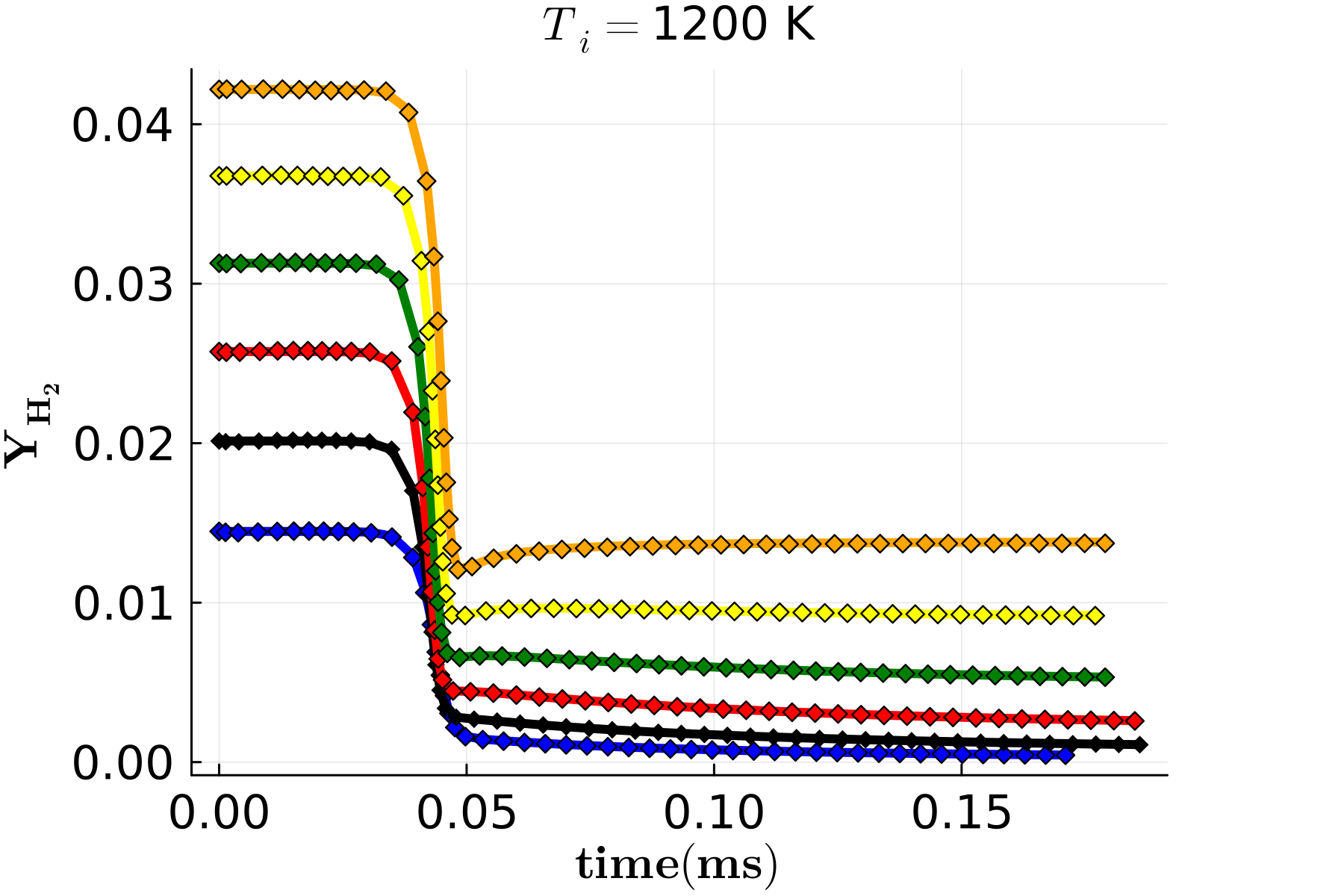}
         \centering
         \centering
         \includegraphics[width=0.32\textwidth]{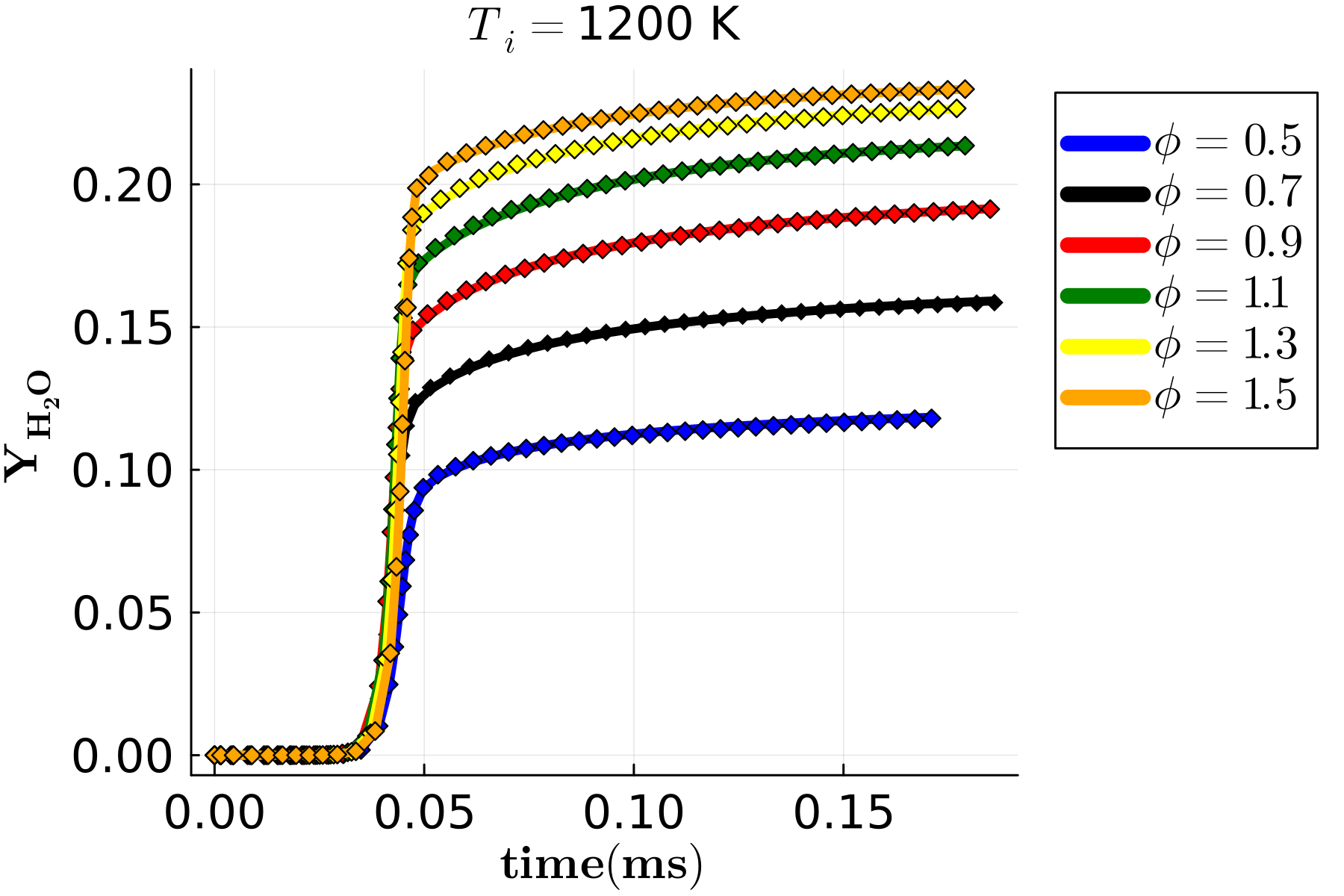}
         \caption{}
     \end{subfigure} 
     \caption[\textwidth]{Temporal predictions from PC-NODE model for temperature and species mass fractions ($Y_{H_2}, Y_{H_2O}$)  at initial temperatures: a) 1000 K, and b) 1200 K, and various equivalence ratios ($\phi$). The solid lines represent the ground truth and markers represent the PC-NODE predictions.}
     \label{fig::profiles}
\end{figure*}

Figure \ref{fig::profiles} shows profiles of the temporal evolution of temperature, mass fraction of one of the reactants ($H_2$), and product mass fraction ($H_2O$) inferred from trained PC-NODE coupled with Cantera for chemical source term computation. The profiles are presented for $T_i = 1000$ K and $1200$ K, and at various equivalence ratios. Overall, very good agreement can be seen between the ground truth (solid lines) and predictions (markers) across different initial temperatures and equivalence ratios from PC-NODE. It is noted that the NODE achieves an inference speedup of approximately 3x over the detailed kinetic mechanism. 

\begin{figure*}
     \centering
         \begin{subfigure}[b]{\textwidth}
             \centering
             \includegraphics[width=0.32\textwidth]{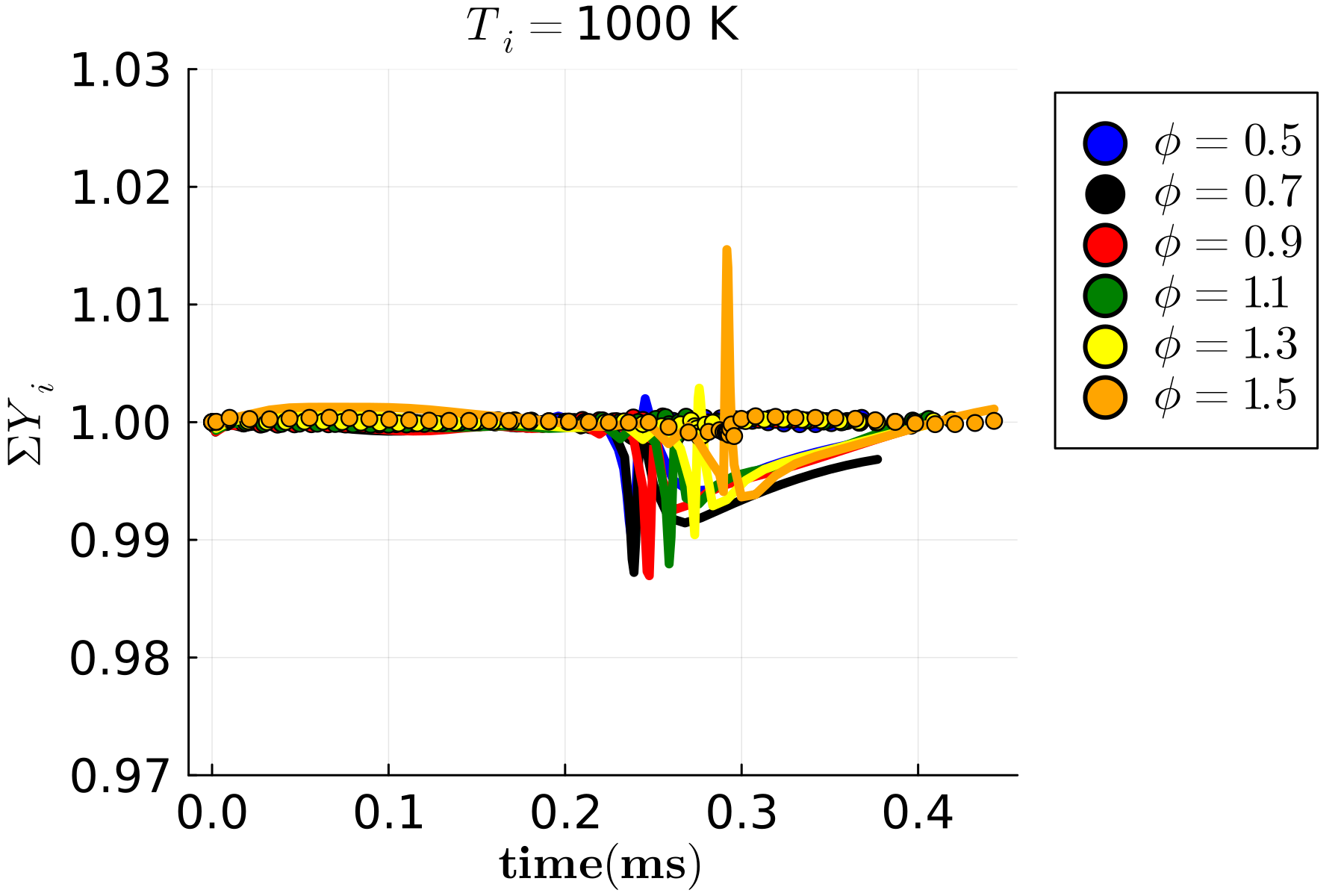}
             \centering
             \includegraphics[width=0.32\textwidth]{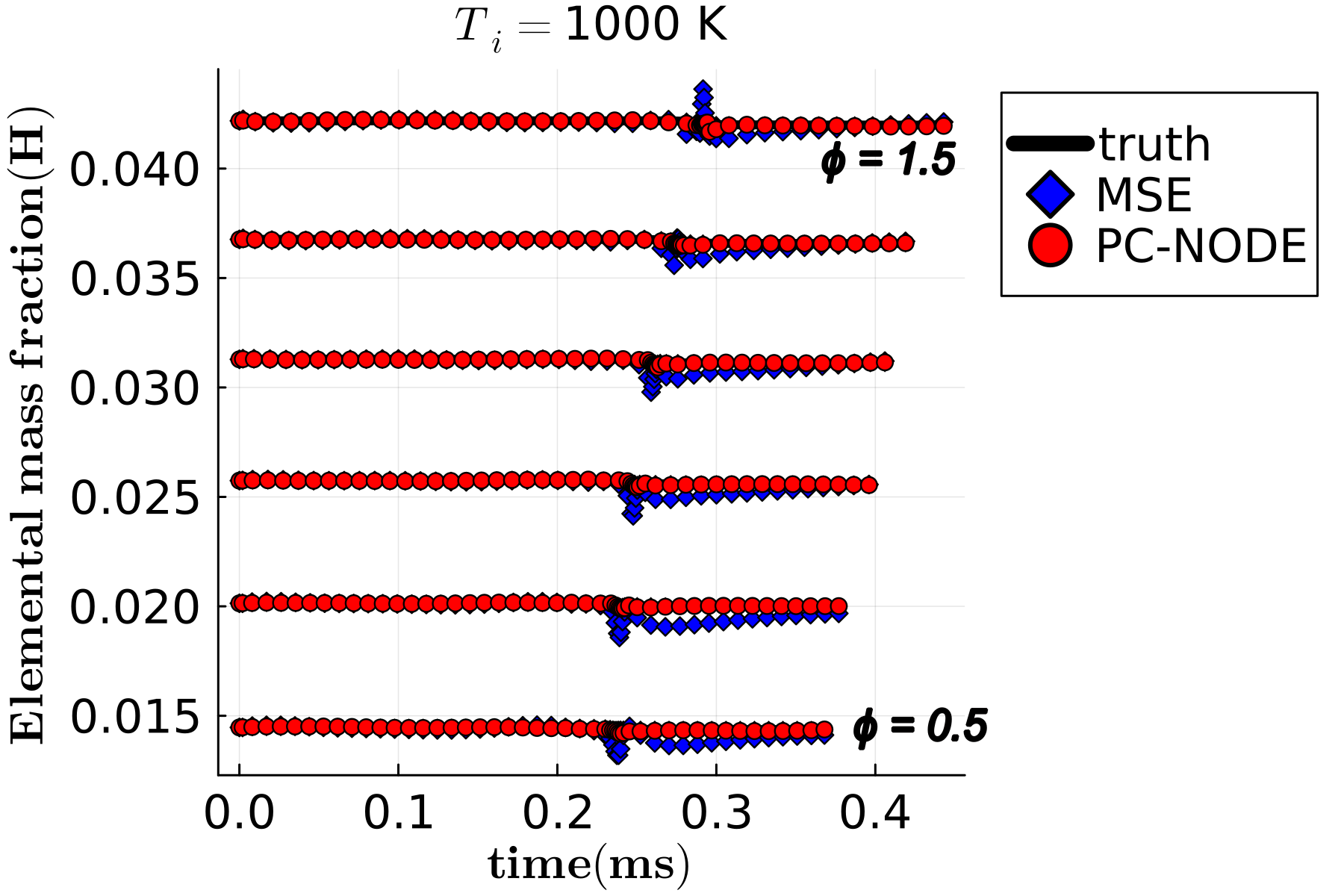}
             \centering
             \includegraphics[width=0.32\textwidth]{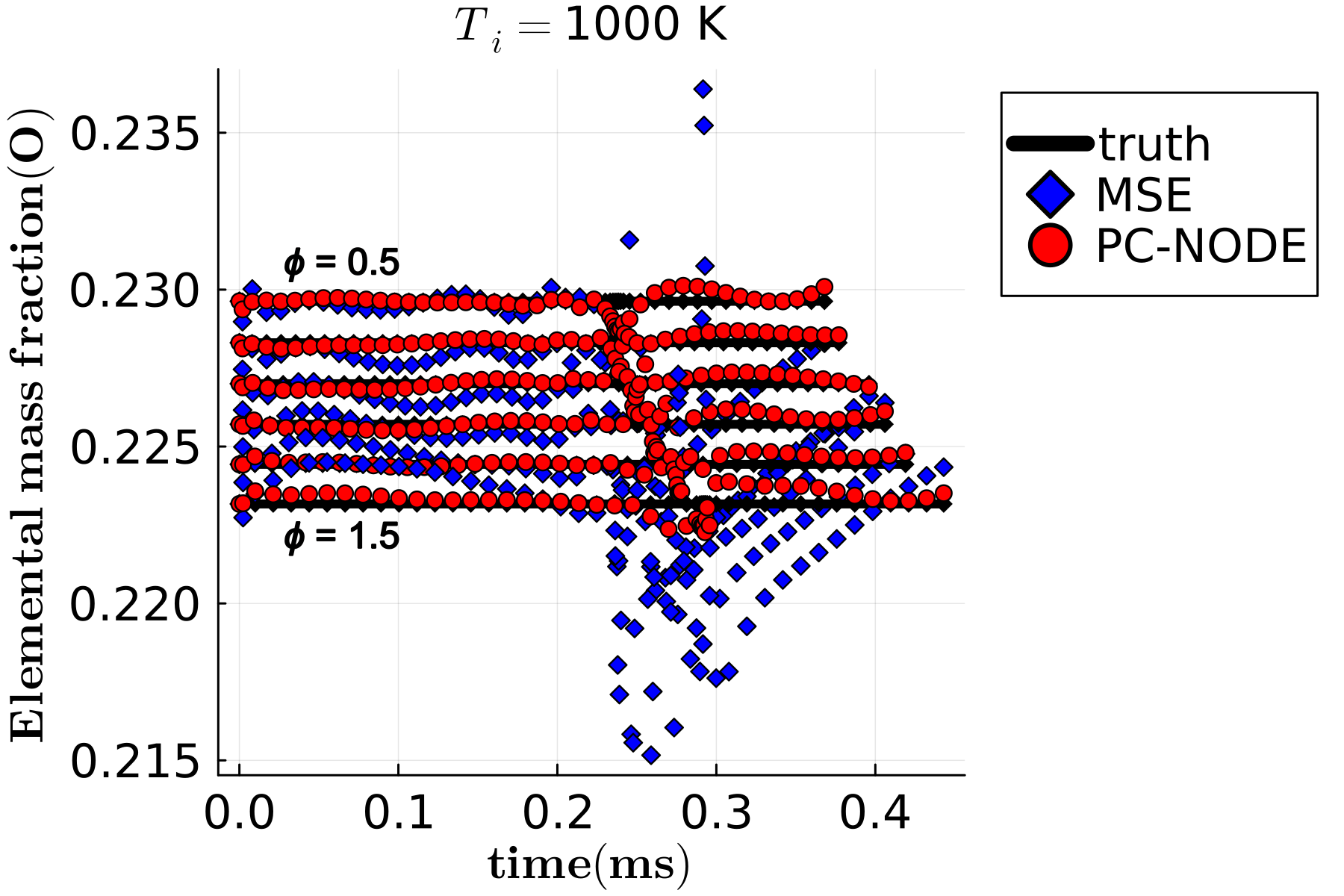}
             \caption{}
          \end{subfigure}
         \vspace{10pt}
          \begin{subfigure}[b]{\textwidth}
             \centering
             \includegraphics[width=0.32\textwidth]{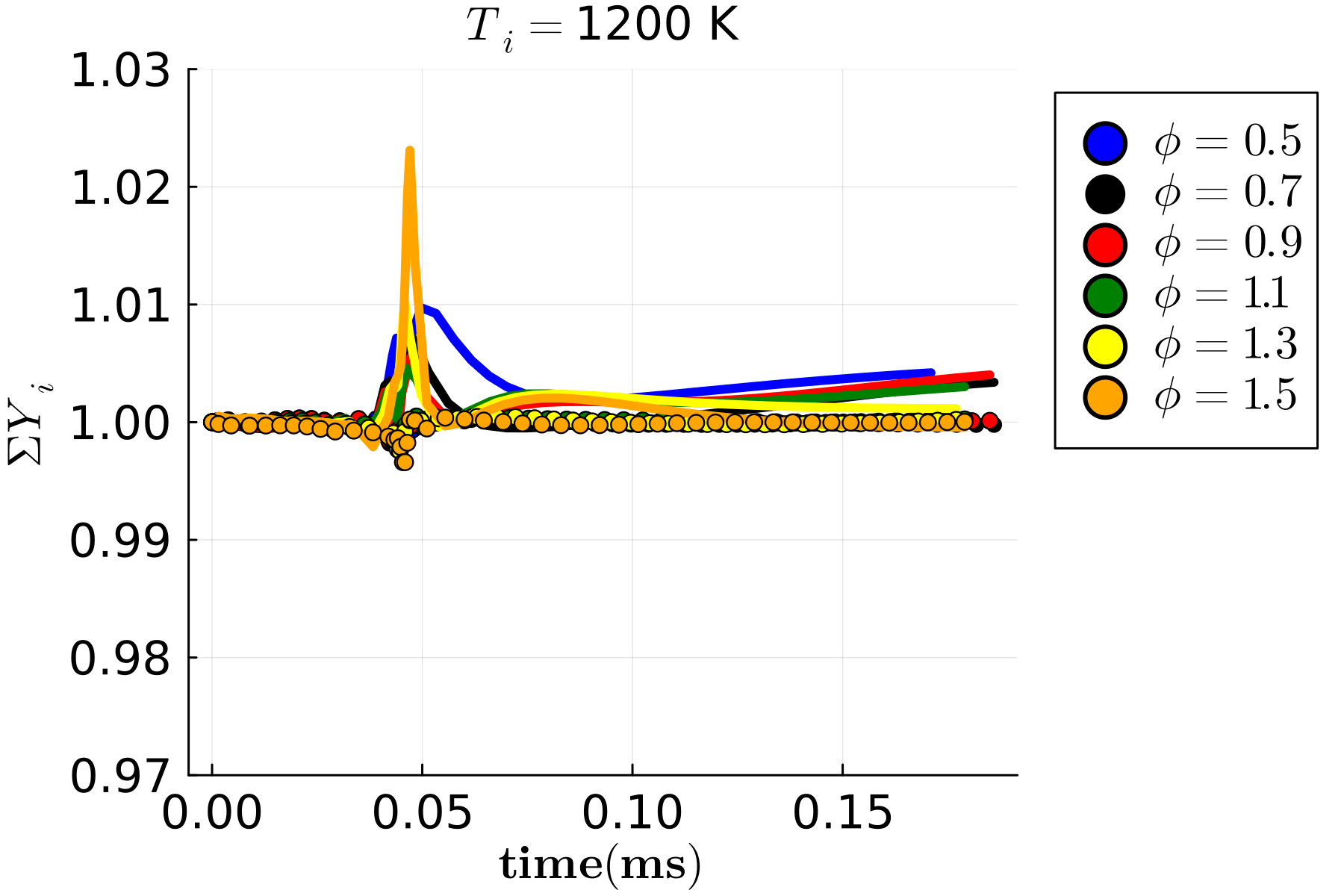}
             \centering
             \includegraphics[width=0.32\textwidth]{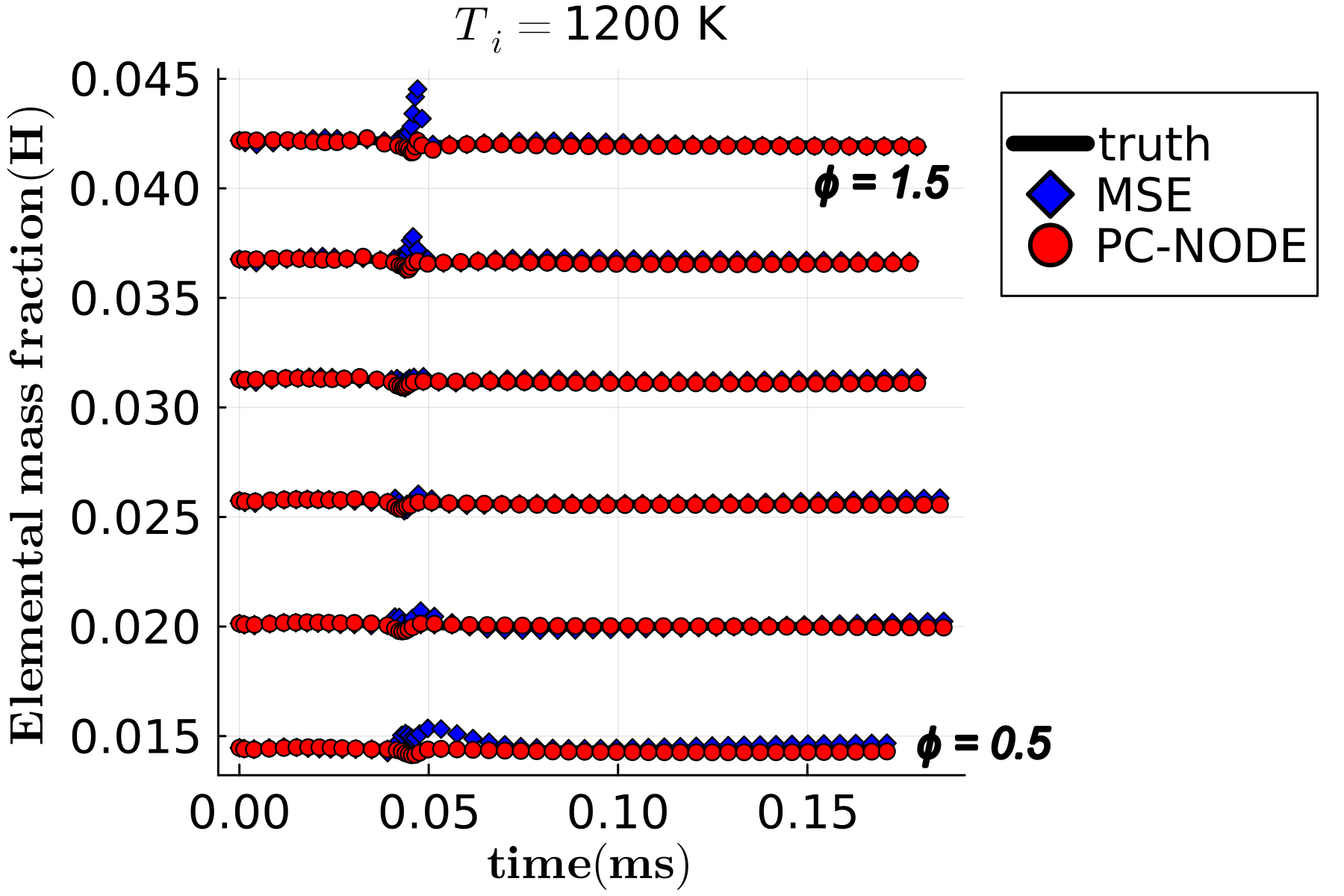}
             \centering
             \includegraphics[width=0.32\textwidth]{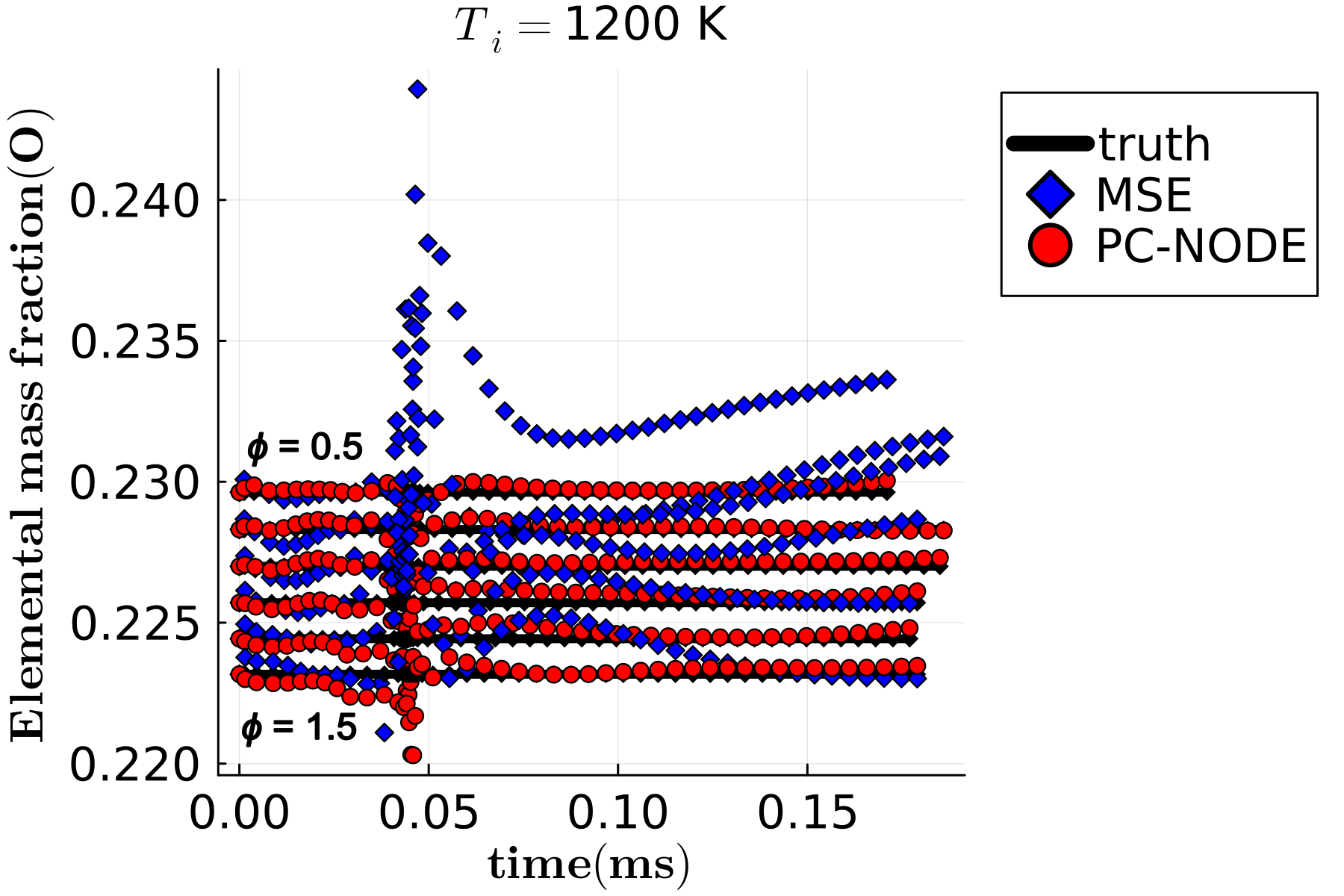}
             \caption{}
          \end{subfigure}

         \caption{Comparison of sum of species mass fractions between MSE trained neuralODE (solid lines) and the physics-constrained neuralODE (markers), and the elemental mass fraction comparison for hydrogen (H) and oxygen (O) at a) $T_i = 1000$~K, and b) $T_i = 1200$~K,  between the cases trained with MSE loss function (blue diamonds) and PC-NODE loss function (red circles). Different equivalence ratios are shown in the same plot.}
          \label{fig::mass_cons_jl}

\end{figure*}

To investigate the effect of loss function constraints on conserved quantities, the total mass, and elemental mass for hydrogen (H) and oxygen (O) are compared. It is seen that the PC-NODE also satisfies the overall species mass conservation significantly better (Fig. \ref{fig::mass_cons_jl}), without an explicit constraint being used for it during training. The ground truth value of sum of species mass fractions is 1, however there is marked deviation from this for the case where only MSE loss function is used. 

Figure \ref{fig::mass_cons_jl} also compares the elemental mass fractions of both hydrogen and oxygen across different initial conditions, for the cases trained with and without the elemental mass constraints. With explicit constraints, elemental mass fraction for both $H$ and $O$ (red markers) in the predicted solution are much better conserved. The case without constraints (blue markers) displays significant deviations, which can result in non-physical solutions when combined with CFD solvers, as they require very low errors with respect to mass conservation laws. Overall, adding physics constraints to neuralODE training helps reduce errors in conserved physical quantities.

\begin{figure}[!h]
     \begin{subfigure}[b]{\linewidth}
         \centering
         \includegraphics[width=0.7\textwidth]{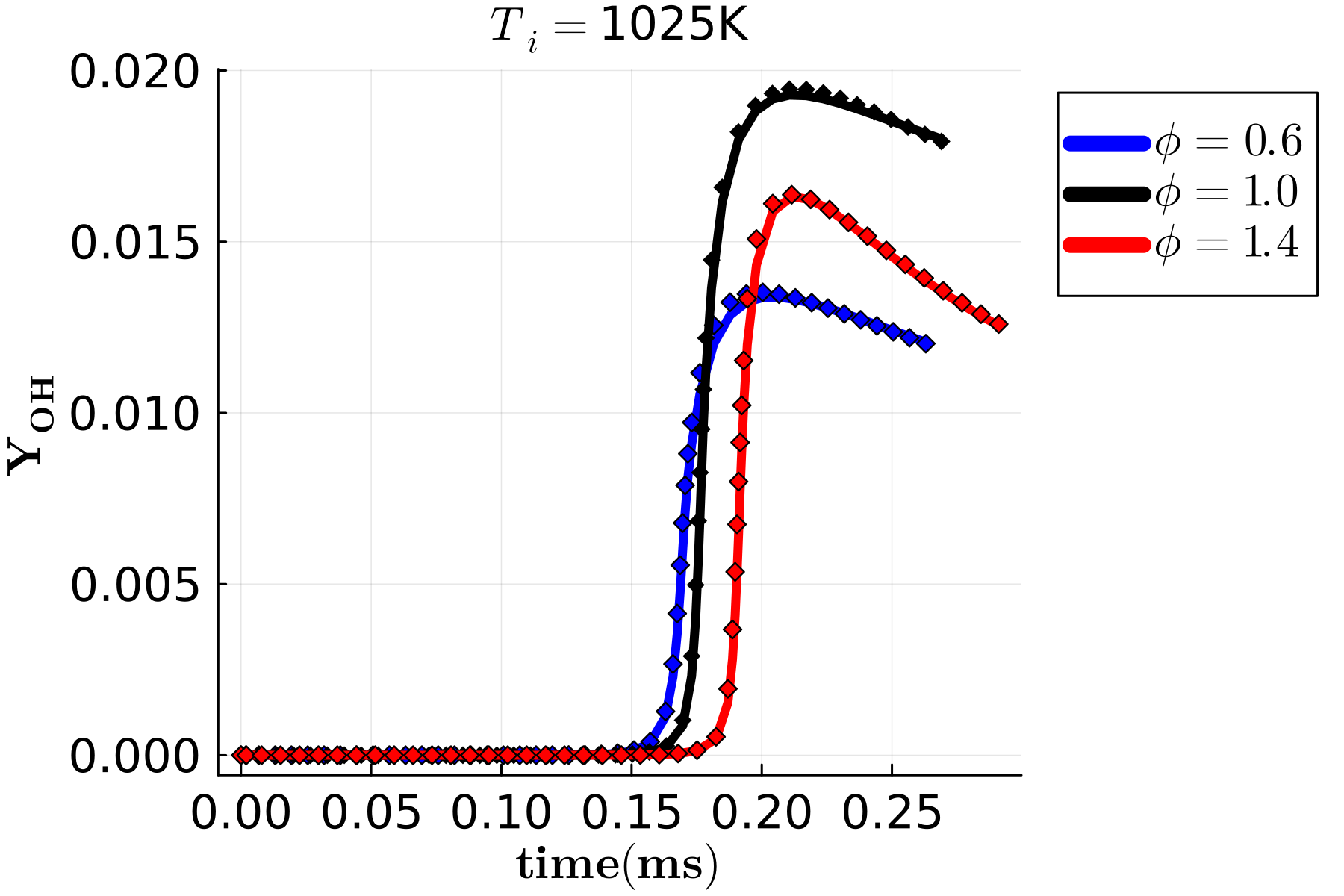}
         \caption{}
     \end{subfigure} \ \
     
    \begin{subfigure}[b]{\linewidth}
         \centering
         \includegraphics[width=0.7\textwidth]{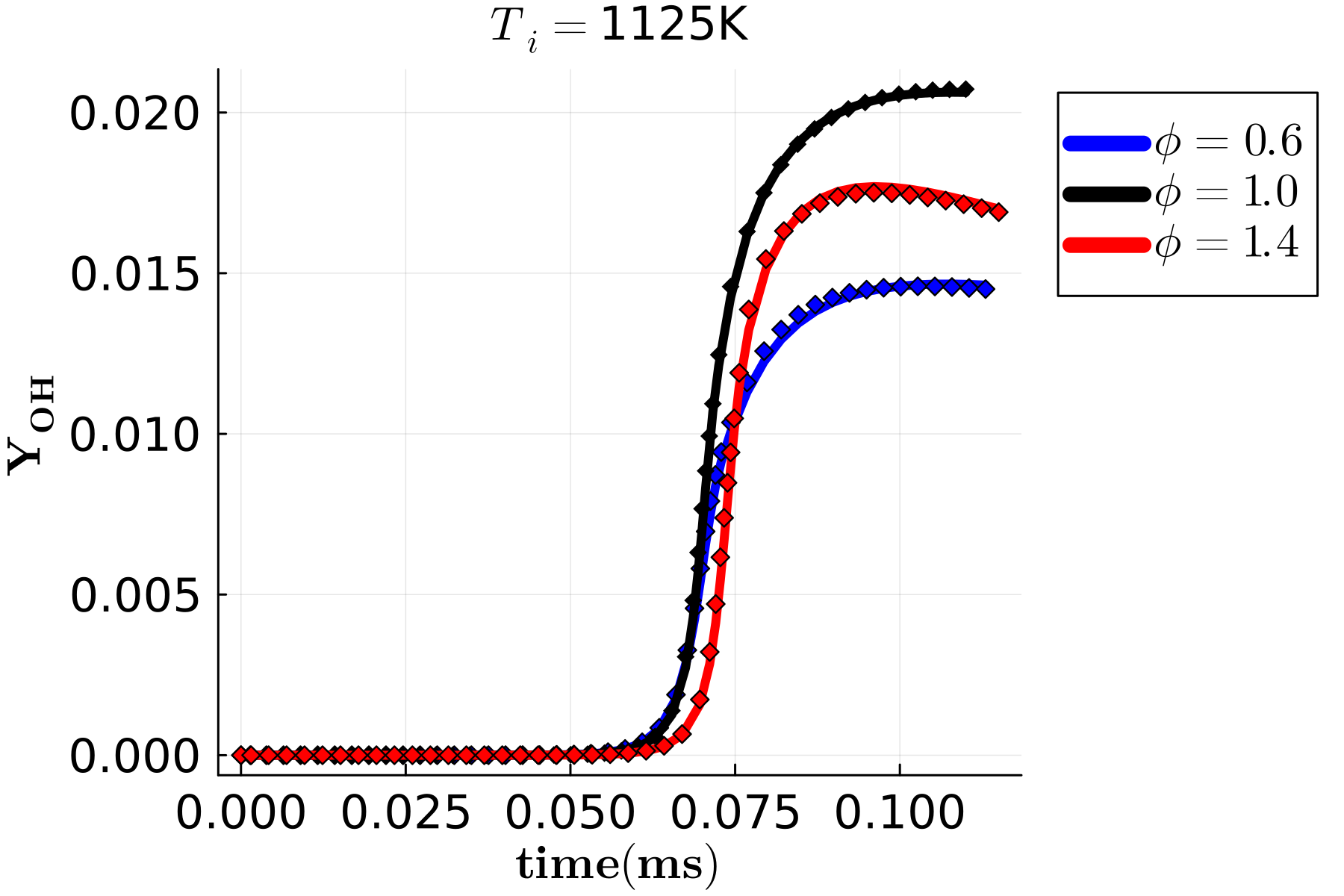}
         \caption{}
     \end{subfigure} 
     \caption[\textwidth]{Temporal predictions from PC-NODE model for $OH$ mass fraction ($Y_{OH}$) for a few initial conditions in the validation set, initial temperatures: a) $T_i =$1025 K, and b) $T_i =$ 1125 K, and various equivalence ratios ($\phi = 0.6,1.0,1.4$). The solid lines represent the ground truth and markers represent the PC-NODE predictions.}
     \label{fig::profiles_valchemnode}
\end{figure}
\begin{figure}[!h]
     \begin{subfigure}[b]{\linewidth}
         \centering
         \includegraphics[width=0.7\textwidth]{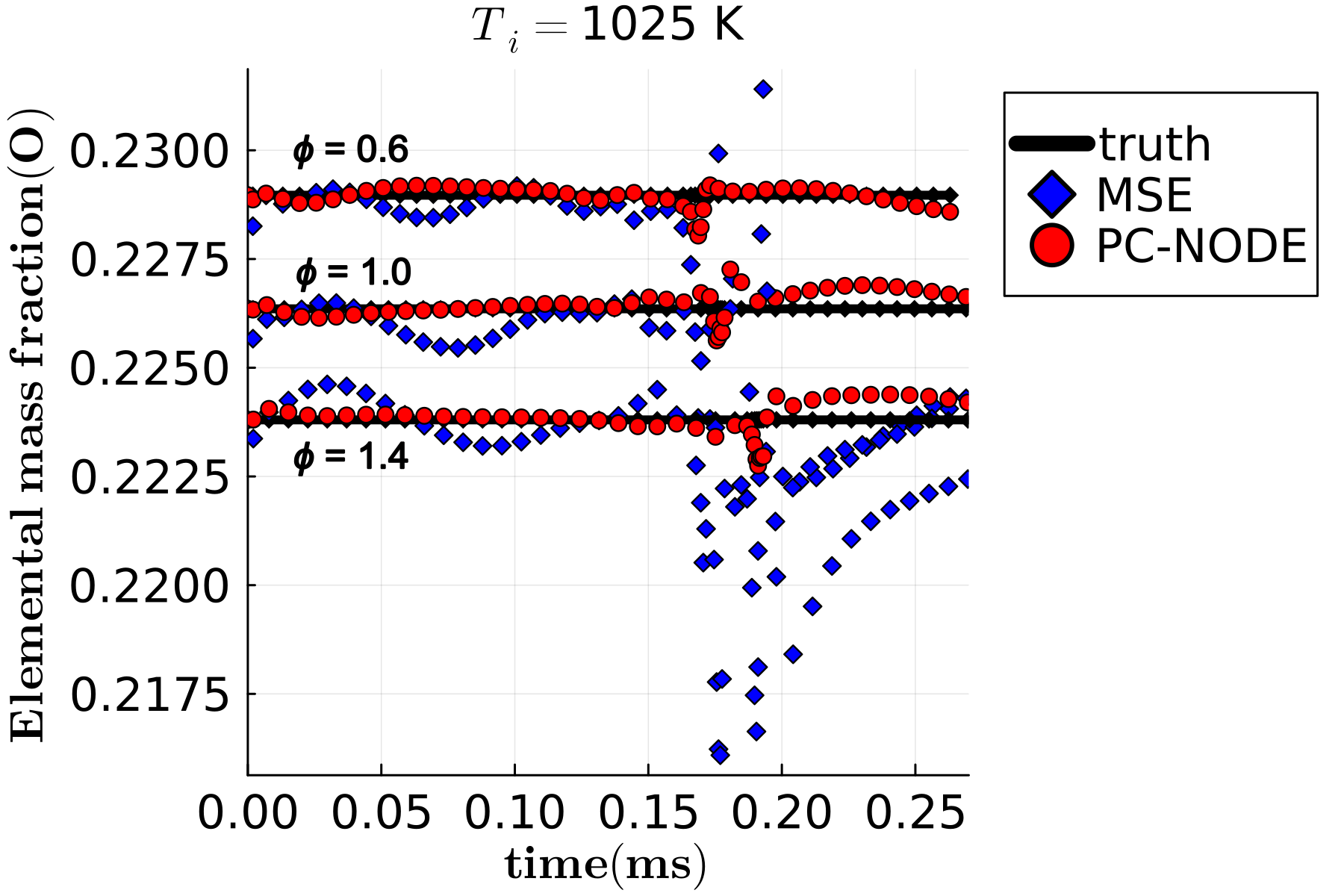}
         \caption{}
     \end{subfigure} \ \
    
    \begin{subfigure}[b]{\linewidth}
         \centering
          \includegraphics[width=0.7\textwidth]{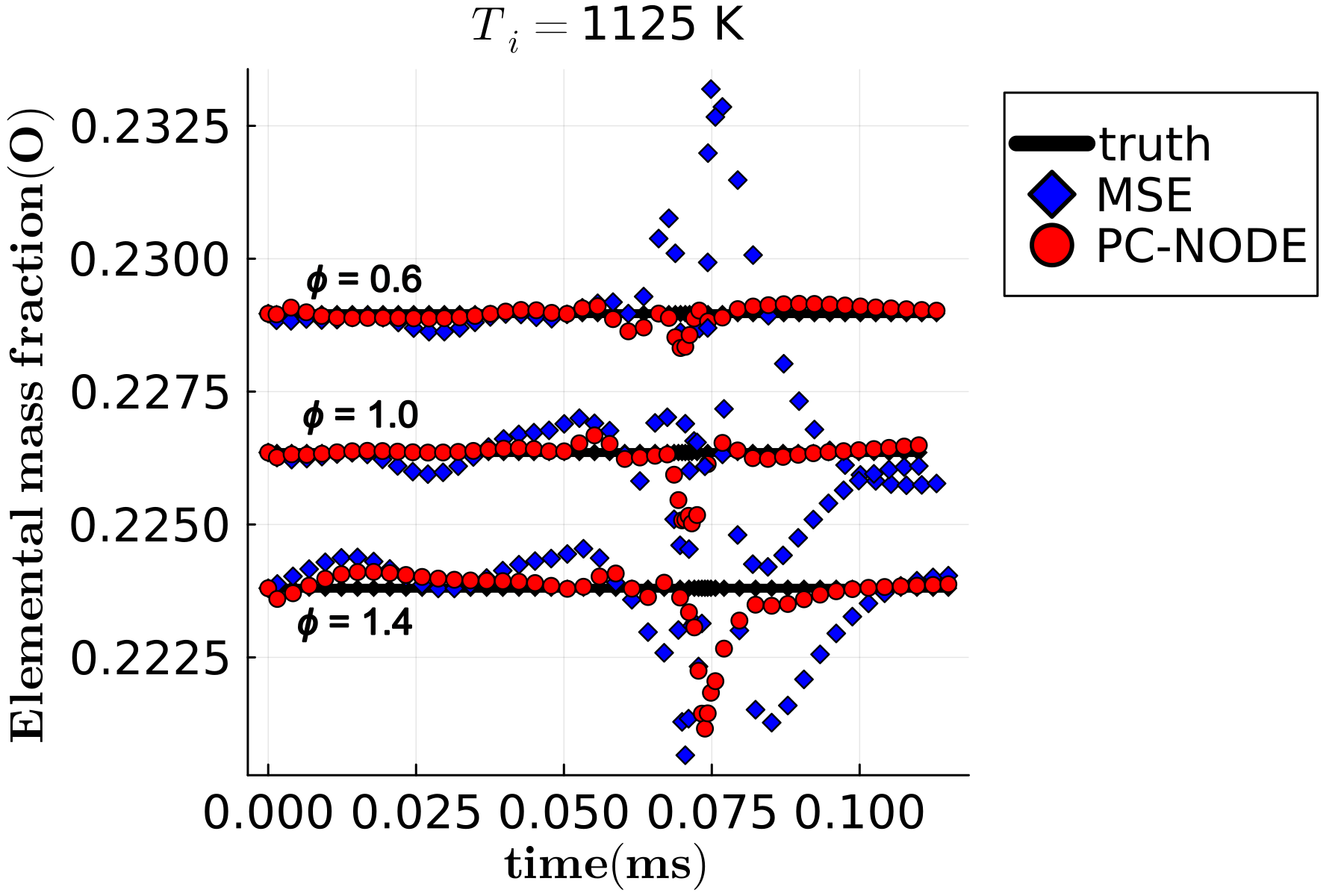}
         \caption{}
     \end{subfigure} 
     \caption[\textwidth]{Comparison of elemental mass fraction of oxygen at initial temperatures: a) $T_i =$ 1025 K, and b) $T_i =$ 1125 K between the cases trained with MSE loss function (blue diamonds) and PC-NODE loss function (red circles). Different equivalence ratios are shown in the same plot.}
     \label{fig::oxycons_val}
\end{figure}

To test the robustness of trained PC-NODE, predictions are made at thermochemical conditions in the validation data set. Three different equivalence ratios ($\phi = 0.6, 1.0, 1.4$) are considered at initial temperatures of $T_i = 1025$ K and 1125 K. Figure \ref{fig::profiles_valchemnode} shows the profiles of the temporal evolution of $OH$ mass fraction ($Y_{OH}$) inferred from PC-NODE. It can be seen that PC-NODE predictions are in excellent agreement with the ground truth data. Further, Figure \ref{fig::oxycons_val} compares the elemental mass fraction of oxygen (O), for the cases trained with and without the elemental mass constraint. Again, the elemental mass fraction (red markers) in the predicted solution is much better conserved for PC-NODE (red) than the case without constraints MSE (blue). 

\subsection{Coupling with CFD solver}\addvspace{10pt}
To demonstrate the implications of coupling the neuralODE-based framework with 3D CFD, the trained neural networks are integrated within CONVERGE CFD \cite{CONVERGE} solver through a user-defined function (UDF). Within CONVERGE UDF, a workflow is set up to read the network weights from text files, formulate the neural network architecture, and perform the forward pass through the network to compute the chemical source terms. These source term predictions from the network are passed to the CONVERGE species solver. 

\begin{figure}[!t]
    \centering
    \includegraphics[width=0.5\linewidth]{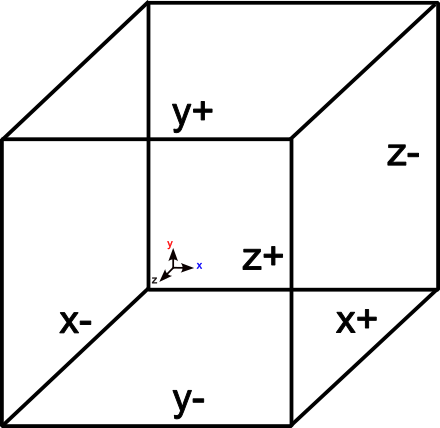}
    \vspace{10pt}
    \caption{Single Cell configuration with the six labeled faces.}
    \label{fig::box}
\end{figure}

To replicate the zero-dimensional constant pressure homogeneous reactor in CONVERGE, a single cubic cell is considered with an edge length, $l = 100 \mu$m, with $x,y,z \in [0, l]$ as shown in Figure \ref{fig::box}. Dirichlet boundary condition ($p = 1$ atm) is applied for pressure at the $x^+$ boundary, while all the other variables (velocity, temperature, species) have zero gradients. On all the other boundaries ($x^-, y^+, y^-, z^+, z^-$), symmetry boundary condition is used for all the primitive variables. The unit cell is initialized with the unburnt gas mixture at specified initial temperature ($T_i$) and equivalence ratio ($\phi$). The first set of studies consist of validation for initial conditions that were used for training the neuralODEs. To this end, both the MSE-trained and physics-constrained NODE (PC-NODE) are integrated within CONVERGE. Tests are carried out for an initial temperature of $T_i = $ 1000 K and equivalence ratios $\phi = 0.5, 1.5$.
 
 \begin{figure*}[!ht]
     \centering
         \begin{subfigure}[b]{\textwidth}
             \centering
             \includegraphics[width=0.32\linewidth]{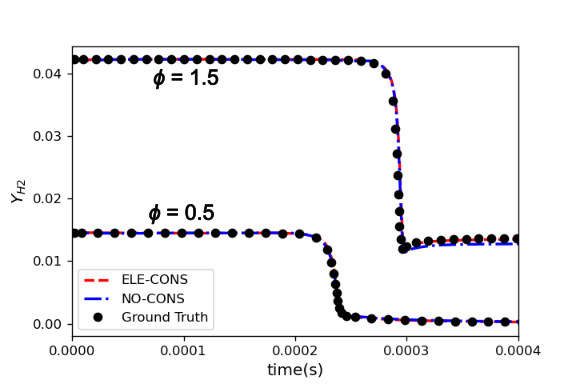}
             \includegraphics[width=0.32\linewidth]{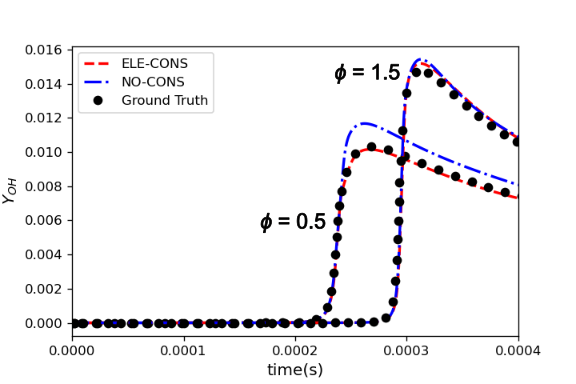}
             \includegraphics[width=0.32\linewidth]{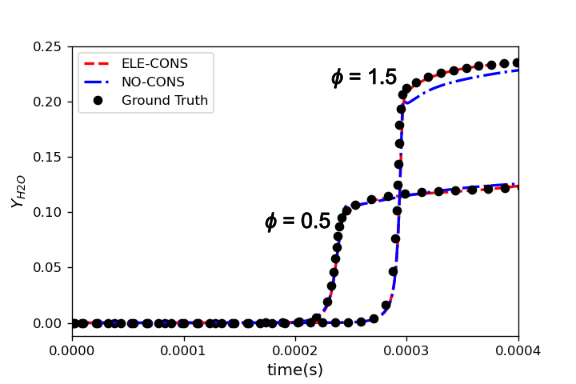}
          \end{subfigure}

         \caption{Temporal predictions for mass fractions ($Y$) of $H_2$, $H_2O$, and $OH$ from neuralODE-integrated CONVERGE simulations at in-sample conditions ($T_i = $ 1000 K and equivalence ratios $\phi = 0.5, 1.5$). The markers represent the ground truth and dashed lines represent predictions. Different equivalence ratios are shown in the same plot. ELE-CONS (red line) is the PC-NODE case, and NO-CONS (blue line) is the MSE-NODE case.}
         \label{fig::conv_plots}
\end{figure*}

\begin{figure}[!h]     
     \centering
     \includegraphics[width=0.7\linewidth]{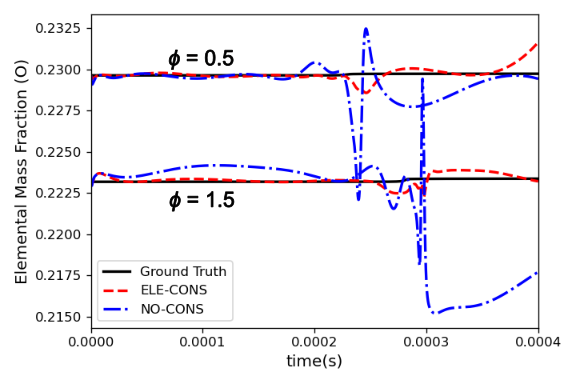}
     \caption[\textwidth]{Comparison of elemental mass fraction of oxygen (O) in neuralODE-integrated CONVERGE simulations, for initial temperature of $T_i = $ 1000 K and equivalence ratios $\phi = 0.5, 1.5$. ELE-CONS (red line) is the PC-NODE case, and NO-CONS (blue line) is the MSE-NODE case.}
     \label{fig::conv_eleOtr}
\end{figure}

Figure \ref{fig::conv_plots} shows the profiles for the temporal evolution of mass fractions of $H_2, O_2, OH$ and $H_2O$ for the two different equivalence ratios, inferred from the network trained without elemental mass constraints (NO-CONS) and the one trained with constraints (ELE-CONS). It can be seen that PC-NODE integrated predictions (ELE-CONS, red line) are in excellent agreement with the ground truth data. The case trained without constraints (NO-CONS, blue line), on the other hand, shows deviations from the ground truth. Further, Fig. \ref{fig::conv_eleOtr} compares the elemental mass fraction of oxygen (O) for these cases. Similar to earlier observations, the elemental mass is better conserved for the PC-NODE cases.  

To further examine generalization capabilities of PC-NODE, the single cell case is simulated at an initial temperature $T_i = $1200 K and equivalence ratios $\phi = 0.4, 1.6$. These conditions were not included in the training set, and are outside the bounds of training data. This helps to test the extrapolation capabilities of the framework. 
 \begin{figure*}[!ht]
     \centering
         \begin{subfigure}[b]{\textwidth}
             \centering
             \includegraphics[width=0.32\textwidth]{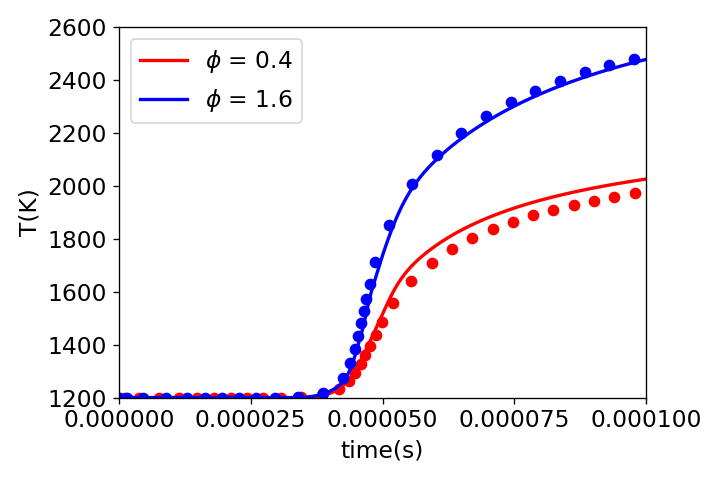}
             \centering
             \includegraphics[width=0.32\textwidth]{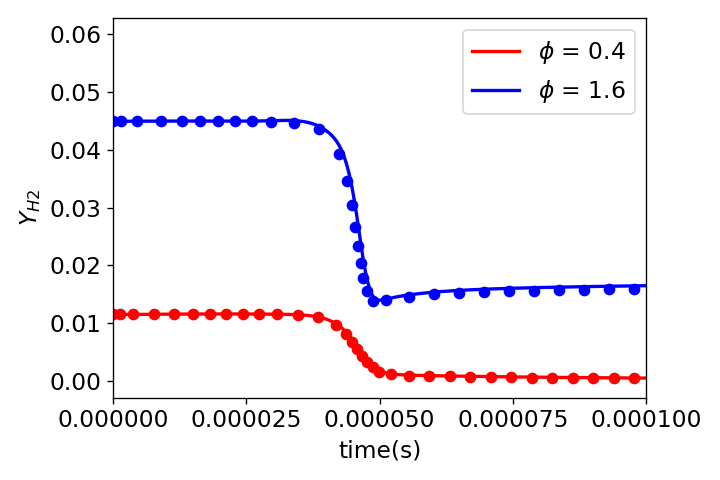}
             \centering
             \includegraphics[width=0.32\textwidth]{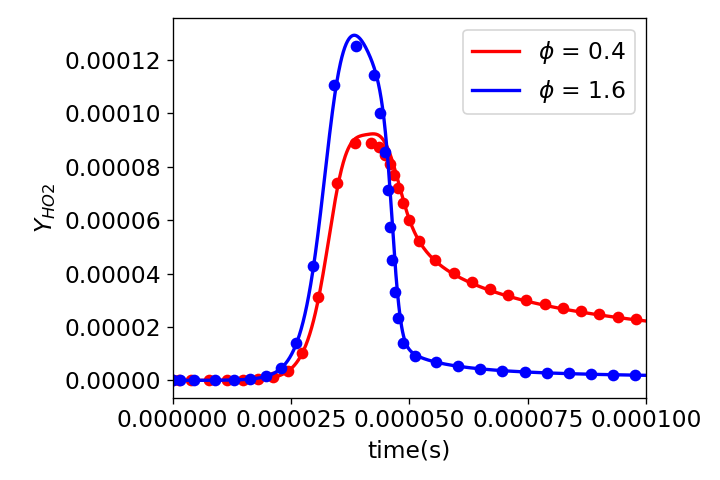}
             \caption{}
          \end{subfigure}
          \begin{subfigure}[b]{\textwidth}
             \centering
             \includegraphics[width=0.32\textwidth]{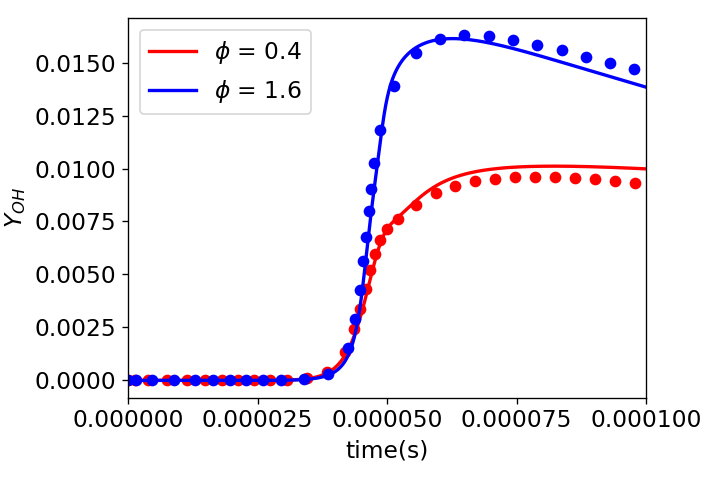}
             \centering
             \includegraphics[width=0.32\textwidth]{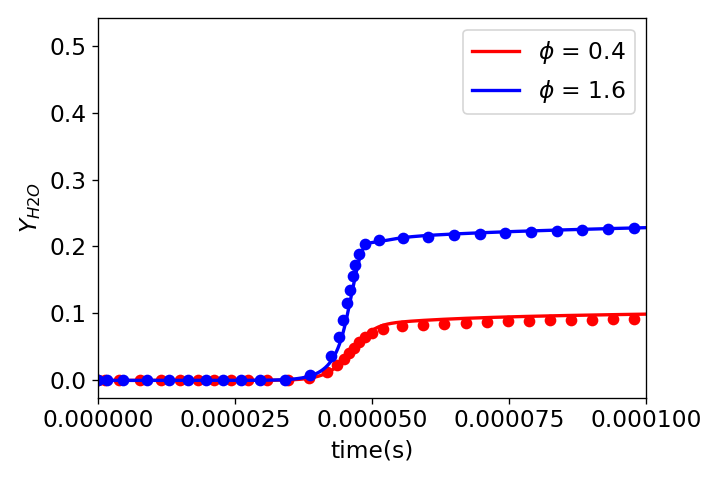}
             \centering
             \includegraphics[width=0.32\textwidth]{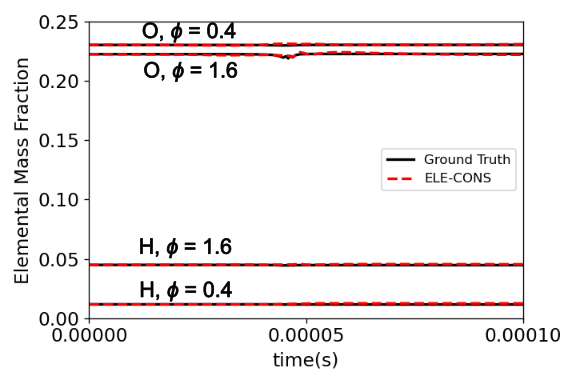}
             \caption{}
          \end{subfigure}

         \caption{Temporal predictions for temperature ($T$) and mass fractions of reactant ($H_2$), product ($H_2O$), radicals ($OH, HO_2$) at out-of-sample conditions ($T_i=$1200K and $\phi = 0.4,1.6$). The solid lines represent the PC-NODE predictions and markers represent the ground truth. Elemental mass fractions for H and O are also plotted. ELE-CONS (red line) is the PC-NODE case.}
         \label{fig::oos_conv}
\end{figure*}
Figure \ref{fig::oos_conv} shows the temporal profile plots for temperature ($T$), and mass fractions of reactant ($H_2$), product ($H_2O$), and radicals ($OH, HO_2$) for these conditions using PC-NODE. The profiles show very good overall agreement with ground truth, with slight mismatch in $Y_{OH}$ and $T$ predictions. Further, Figure \ref{fig::oos_conv} also shows the PC-NODE predicted elemental mass fractions of hydrogen (H) and oxygen (O) for these conditions, which are in excellent agreement with ground truth.

\section{Conclusions}\addvspace{10pt}
In this work, a novel physics-constrained neuralODE (PC-NODE) framework for stiff chemical kinetics was introduced by incorporating elemental mass conservation constraints directly into the loss function during training of the data-driven model. This ensures that both the total mass and elemental species mass are conserved. Proof-of-concept studies were performed for homogeneous autoignition of hydrogen-air mixture
over a range of composition and thermodynamic conditions at constant pressure. The results showed that the proposed enhancement not only improves the  physical consistency with respect to mass conservation criteria but also improves training efficiency.

The trained NeuralODE framework was further integrated with CONVERGE CFD solver, and demonstration studies were conducted for hydrogen-air constant-pressure homogeneous autoignition in a 3D unit cell configuration. The coupled framework showed excellent agreement with the detailed chemical kinetic mechanism, and the prediction from PC-NODE exhibited significantly lower errors in temporal evolution of thermochemical scalars and much smaller deviation from mass conservation constraints compared to purely data-driven NODE approach. The PC-NODE framework also demonstrated very good generalizabition capability, at thermochemical conditions outside the training range. In future work, the PC-NODE framework will be extended to larger kinetic mechanisms and demonstrated for reacting CFD modeling of combustion systems.

\acknowledgement{Declaration of competing interest}\addvspace{10pt}
The authors declare that they have no known competing financial interests or personal relationships that could have appeared to influence the work reported in this paper.
\acknowledgement{Acknowledgments} \addvspace{10pt}
The submitted manuscript has been created by UChicago Argonne, LLC, Operator of Argonne National Laboratory
(Argonne). Argonne, a U.S. Department of Energy Office of Science laboratory, is operated under Contract No.
DEAC02-06CH11357. The U.S. Government retains for itself, and others acting on its behalf, a paid-up nonexclusive,
irrevocable worldwide license in said article to reproduce, prepare derivative works, distribute copies to the public, and
perform publicly and display publicly, by or on behalf of the Government. The research work was funded by the DOE Fossil Energy and Carbon Management (FECM) office through the Technology Commercialization Fund (TCF) program. The authors thank Dr. Shuaishuai Liu from Convergent Science, Inc. for helpful discussions on coupling the neuralODE framework with CONVERGE CFD solver. The authors would also like to acknowledge the computing core hours available through the Bebop and Swing clusters provided by the Laboratory Computing Resource Center (LCRC) at Argonne National Laboratory.


 \footnotesize
 \baselineskip 9pt


\bibliographystyle{pci}
\bibliography{PCI_LaTeX}


\newpage

\small
\baselineskip 10pt



\end{document}